# EXPERIENCE CAPTURE IN SHIPBUILDING THROUGH COMPUTER APPLICATIONS AND NEURAL NETWORKS

**S. U. Sangeet**, **K. Sivaprasad, Yashwant R. Kamath,** Department of Ship Technology, Cochin University of Science and Technology, Kerala, India

**ABSTRACT**

It has always been a severe loss for any establishment when an experienced hand retires or moves to another firm. The specific details of what his job/position entails will always make the work more efficient. To curtail such losses, it is possible to implement a system that takes input from a new employee regarding the challenges he/she is facing and match it to a previous occurrence where someone else held his/her chair. This system could be made possible with input through the ages from the array of individuals who managed that particular job and processing this data through a neural network that recognizes the pattern. The paper is based on data collected from traditional wooden dhow builders and some of the modern day unconventional shipyards. Since the requirements for successful implementation in such scenarios seems too steep at the moment, an alternate approach has been suggested by implementation through the design processes across multiple shipyards. The process entails the traditional value passed down through generations regarding a particular profession and analysis has been done regarding how this knowledge/experience can be captured and preserved for future generations to work upon. A series of tools including SharePoint, MATLAB, and some similar software working in tandem can be used for the design of the same. This research will provide valuable insight as to how information sharing can be applied through generations for effective application of production capabilities.

## 1. NOMENCLATURE

| | |
|---|---|
| ANN | Artificial Neural Network |
| EULA | End User License Agreement |
| GUI | Graphical User Interface |
| IACS | International Association of Classification Societies |
| KBMS | Knowledge Base Management System |
| NPO | Non-profit Organisation |

## 2. INTRODUCTION

Since the advent of ships, there have been accidents at sea. The root causes for all those unfortunate events were lack of attention to detail in both maintenance and newly built ships. Human errors were the main reason for almost 80-85% of maritime accidents [2]. The entities that are responsible to supervise and enforce attention to detail are the classification societies that ensure that shipbuilding conform to certain acceptable standards. It would be a huge leap if we could ensure better quality and increase production efficiency with a single tool. This paper finally provides a proposal that could accomplish just that.

## 3. METHODOLOGY

Two approaches were considered based on reach and economic viability. Since computational power has been keeping up with the fast advancing requirements for the past decade, this method has acquired relevance only recently. Therefore it is only justified that a case study be conducted to verify the applicability of such a design. Analysis was done through case studies that finally provided an idea as to how it would help or harm the current scenario. Therefore this paper projects the results of a case study.

## 4. CASE STUDY

4.1    Case 1- Implementation in Production

4.1 (a)  Applicability in a Shipyard

It is possible to implement a database development programme to serve as a base for creating a knowledge system that is preformatted as per the requirements for an artificial neural network that can perform deep learning. Deep Learning recognises intricate patterns in huge databases that has varying levels of hierarchy through a backpropagation algorithm and suggests how a machine should re-compute each layer using different parameters [1]. With this, a probable neural network that applies here can calculate the extent of dependency between databases generated from different hierarchical positions. In the present day scenario, this applies to the different levels of communication between the design, production and outfit departments in a new build shipyard [3]. The most viable way to implement such a database development programme would be in the production department where decisions are made every day as to what the next step should be in assembly, transportation and welding in general. A lot of time can be saved by creating a comprehensive timeline where every department can work in tandem without disrupting another's work. User friendly GUI's can further alleviate the pains associated with training the working professionals to log their work into the database every day. Microsoft SharePoint is a platform where shipyard workers can log their everyday decisions and subsequent effects [4]. So the prime requirement from our solution is anomaly detection in the logs created through Microsoft SharePoint. As shown in Figure 1, there can be different types of anomalies depending on the type of ship being built. Since shipbuilding is one off kind, there can be point anomalies like the one in Figure1-(B) which could simply mean a different type of outfitting as requested by the owner that calls for a specific type of welding in a location or timeframe which is not found among the data from other shipbuilding activities. There can also be collective anomalies like the one shown in the Figure 1-(C) which could be due to an entirely different kind of ship being built which involves entirely different procedures in unit assembly, sub-assembly, welding, outfitting and launching [5]. Together, they account for a collection of anomalies that are grouped together. If we assume that our observations are based on a stochastic model, statistical analysis can be used to forecast further data points with occasional anomalies [5]. Figure 1-(A) represents a typical new build job with no anomalies in any of the different processes involved in the build process.

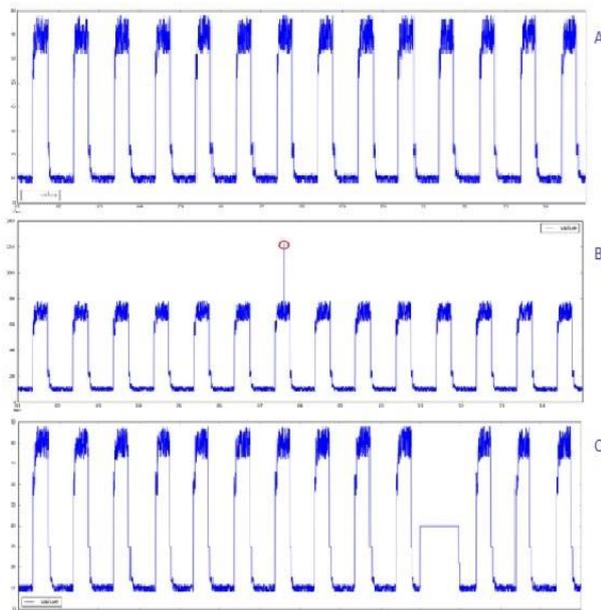

Figure 1- Statistical representation of data points collected. (A): Normal Distribution, (B): Anomaly in single value of data, (C): Collective Anomalies

Simply put, after such a neural network has been subjected to deep learning and is properly trained, the ANN can forecast the entire timeline for the shipbuilding process including many circumstances that may be overlooked generally with solutions to get around those difficulties. This could be a huge impact on the production scenario in shipyards and could easily boost production efficiency drastically. At the same time it could save a lot of time and effort for the planning team. Such a system in its entirety could be attributed the name, KBMS, a knowledge based management system [6]. A series of steps necessary for the implementation of such a system in a shipyard is detailed below in chronological order:

- Training of Personnel
- Infrastructure
- HR allocation
- Database Review Team
- Database Incubation
- Periodical analysis of data points
- Minimum Database Collected
- Integration of Neural Network into the Database System
- Recognising Parameters
- Setting up Input/output
- Defining Neural Network Dynamics
- Training the network
- Pattern Recognition
- Suggestion of solutions (output**)**.

The above process will require at least a decade to properly train the network to better analyse and provide solutions for problems. Since this follows a statistical approach, the only way to improve accuracy is to add more data points provided the network definition is sound. A program with such a long term goal can only be undertaken by a body that has several shipyards working around the world that can provide different databases to efficiently train the network and make it capable of dealing with ships of all types and sizes. Any other shipyard would find it wasteful both economically and in human resources. Even here, classification societies could together (IACS) create a database for the same using data from shipyards all around the world.

4.1 (b) Applicability in Traditional Shipbuilding

Traditional wooden dhow construction carried out in Kerala, India is the perfect example of knowledge capture. The head shipwright also known as 'Moothasari' in local language trains an apprentice as his successor. The apprentice works with the Head shipwright for more than half of his mentor's working tenure until the mentor deems his successor worthy to take over. Therefore this would be the perfect platform to implement the deep learning concept and create an artificial neural network that can later on become the apprentice's personal troubleshooting tool. In the said context, it is still uncertain whether this can be implemented immediately without an overseeing body that can provide technical assistance. Hence this approach requires more resources including financial funding. Therefore this method is not feasible presently but could be very useful in the future provided classification societies can take a more active role in analysing and classifying traditional wooden dhows.

4.1 (c)  Applicability in a NPO

A KBMS may contribute differently to different shipyards around the world that predominantly deals in certain types or sizes of ships. Hence it is more sensible to create a global model that can be trained with data from shipyards around the world, capable of dealing with huge amounts of data on all varieties of ocean going vessels. Even though this might require several teraflops or more in computing power, it will certainly yield better efficiency in production and further improve communication of practices among shipyards globally. It is possible to set up a Non-profit organisation (NPO) that is funded and run by classification societies and other institutions like RINA (Royal Institution of Naval Architects). Such an organisation can monitor and provide a common platform for various shipbuilding activities to work in tandem under an array of strict quality control systems that can ensure safety of the vessels at sea while decreasing the production time of every individual ship. This will greatly reduce the time required to train an ANN to comply with all the required norms in shipbuilding. It also adds to the fact that having a great many number of databases to include in the knowledge base will increase the accuracy of all predictions done by the network tremendously. A common data entry form can be provided to all shipyards to log their data and problems faced. Instant troubleshooting will be available to all the employees through real time feedback from the network as well as other working individuals of the same position in a different shipyard. All amendments made by the IMO, IACS and other authorities can be brought to practise with ease and by spending lesser resources through simple parameter manipulation. In an age where macrocosms of computers will become the tools used to govern industries at large, such a tool would be the perfect ground to build upon. It remains to be seen whether a technology will arise that can further simplify the process in the very less timeframe that this approach needs. In event of such a discovery, this method will lose its relevance and applicability.

4.1 (d)  Conclusion of Case 1

In all three methods elaborated on above, it was apparent that applying such a tool successfully requires a number of resources both financially and in man power. Hence, we came to the conclusion that it would be easier to implement the tool at the grassroots level. That is in the design department. Already the shipbuilding industry has become very progressive by developing countless comprehensive software that can deal in both design and production. Hence the second case studies the effects and applicability of an ANN in design software.

4.2     Case 2- Implementation in Design

With time, ship design has aged from full scale drawings by master loftsmen to complete three dimensional renderings of ship models with computerised structural analysis using advanced finite element analysis.

It is now possible to simulate almost all conditions that the vessel might face during its lifetime on desktop based computers. As shown in Figure 2, all parts including structure, tanks, plumbing, outfitting etc. can be simulated along with the hull for exact real time data replication. Every aspect of the shipbuilding process from the stockyard to the fully outfitted ship at sea can now be designed, planned and constructed using comprehensive software tools that provide naval architects with everything they need to complete the detailed design. Simulations can even tell us what might happen if we start using some new technology, or different equipment mainly focusing on the cost of operation and strict adherence to build timeline [7]. This field is still evolving. All software are now being equipped with user-friendly, interactive, and easy to learn graphical user interfaces that simplify the process for the working personnel in a shipyard's design department. It would be very easy to train an ANN with deep learning using data from widely used ship design software. Since the same software will be used in shipyards all around the world, there will be numerous data points to serve as input while training the neural network. Hence prediction accuracy will be very high. Since classification societies are already taking an active role in promoting ship design software to incorporate eco friendly building techniques as well as the design of hulls, bows and propellers with higher efficiencies, it will not require a great deal of effort to include a knowledge acquiring system in them. Due to the large user base that such software has in the industry, we can train the ANN in a very short span. Qualitative and quantitative data from over a hundred shipyards can easily train the ANN for life. Such a venture can be tailored to perfection by IACS. Similar to the archives of ship data that some of the classification societies now maintain, a knowledge base that supplements the archives with the details of the building process will easily suffice when collected from a large number of shipyards. Once so many data points are amassed, the network will be able to forecast data on a variety of subjects covering all aspects of the design process.

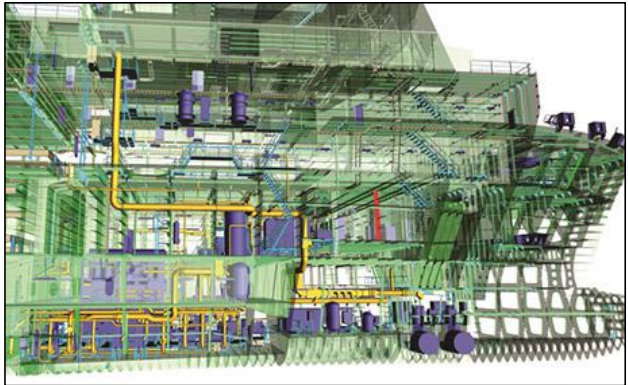

Figure 2- Typical schematic rendered on a design tool [8]

Once it is successfully implemented in the design process, it can be extended to the production stage through class surveyors who supervise the production process. Information can be logged through the small tablet computers that a surveyor takes with him. Such a device can reduce the number of items that he has to carry around as it is an exceptionally task while accessing areas like the double bottoms and wing tanks for survey. If everything starts to be logged electronically starting now, the database we build will be most useful for developing better ways of accomplishing tasks that are otherwise hard to do.

4.2 (a)  Sample Case Study in Suggestion Output

Figure 3 shows a sample data entry form that would be used by the designers in the shipyard to train the ANN for future suggestions. In future when the network is capable of extending beyond the design department, it will need both qualitative and quantitative data from more than a hundred shipyards. For example, a welder in his right mind-set can do his work perfectly while a disturbed one cannot. In everyday shipbuilding there are numerous welded joints being created per day. If a welder is careless in his work, the whole region in the newly built ship might become weak and prone to fracture. In a scenario where the ANN extends its reach into production, it can be monitored whether workers are over exerted or if someone is not good at their jobs. This can improve quality assurance at every shipyard. Moreover different problems faced by different shipyards might have a single and easy solution. Such instances can be easily identified and rectified.

Figure 3- A GUI for Initial Database Creation/ Output

## 5. CONCLUSION

In short, it can be concluded that the easiest and most economical way to start the process of building a KBMS that employs an ANN capable of deep learning is through the premium design software that are widely used. Once a legal EULA is drawn that safeguards the commercial interests of all the users, data can be used to build a knowledge base that will be useful for future generations to build upon. It is a fact that shipbuilding practises have remained more or less the same save for some advanced tools that simplify the process. There have been no radical technological advancements that questioned the relevance of conventionally modelled ships. Hence it is paramount that a digital database for the same be developed that encases all the core values of traditional and present day shipbuilding.

## 6. ACKNOWLEDGEMENTS

The work on this paper started amidst a pandemonium of other responsibilities and ended the same way. Therefore, first and foremost I would like to thank my family who provided valuable advice, support and motivation and so became the prime mover in this venture. This paper would have been impossible without the invaluable help and advice provided by my teacher and mentor, Dr. K Sivaprasad.

## 8. AUTHORS BIOGRAPHY

**K. Sivaprasad** currently holds the position of Associate Professor at Department of Ship Technology, Cochin University of Science and Technology, Kerala, India. His area of specialization is in Ship Production and Ship Repair and so handles relevant classes for the B.Tech Program and is also a Research Guide. He has more than 35 published papers to his name and has more than 20 years of experience in teaching. He has also worked as a Senior Naval Architect at Mazagon Docks Ltd., India.

**S. U. Sangeet** has just completed his bachelor's degree in Naval Architecture and Shipbuilding from Department of Ship Technology, Cochin University of Science and Technology, Kerala, India.

**Yashwant R. Kamath** currently a student of B.Tech in Naval Architecture and Shipbuilding at Department of Ship Technology, Cochin University of Science and Technology, Kerala, India.